\documentclass{article}

     \usepackage[preprint]{neurips_2020}

\usepackage{graphicx}
\usepackage{multirow}
\usepackage{lscape}

\usepackage{lineno}

\usepackage[utf8]{inputenc} 
\usepackage[T1]{fontenc}    
\usepackage{hyperref}       
\usepackage{url}            
\usepackage{booktabs}       
\usepackage{amsfonts}       
\usepackage{nicefrac}       
\usepackage{microtype}      

\usepackage[utf8]{inputenc}
\usepackage{amsthm}
\usepackage{amsmath}
\usepackage{amssymb}
\usepackage{natbib}

\usepackage{bm}
\usepackage[toc,page]{appendix}
\usepackage[algoruled, longend, boxed]{algorithm2e}
\usepackage{multirow}
\usepackage{mathrsfs}
\usepackage{float}
\usepackage{subfigure}
\usepackage{siunitx}
\usepackage{xcolor}

\nolinenumbers

\newcommand{\mtheta}{\bm{\mathit{\theta}}}
\newcommand{\momega}{\bm{\mathit{\omega}}}

\newcommand{\mbx}{\mathbf{x}}
\newcommand{\mby}{\mathbf{y}}

\newcommand{\mbX}{\mathbf{X}}

\newcommand{\mathE}{\mathbb{E}}
\SetKwInput{KwInput}{Input}                
\SetKwInput{KwOutput}{Output}              
\SetKwInput{KWInitialization}{Initialization}                
\SetKwInput{KWReturn}{Return}                

\title{Expectation-Maximization Regularized Deep Learning for Weakly Supervised Tumor Segmentation for Glioblastoma}

\begin{document}

\bibliographystyle{model2-names}
\author{%
  Chao Li \\
  Department of Clinical Neurosciences \\
  University of Cambridge\\
  Shanghai General Hospital\\
  Shanghai Jiao Tong University\\
    \And
 
  Wenjian Huang\thanks{Equal contribution}
  \\Department of Computer Science and Engineering\\
  Southern University of Science and Technology \\

  \And
  Xi Chen \\  
  Department of Computer Science\\
  University of Bath \\
  \And
  Yiran Wei \\
  Department of Clinical Neurosciences \\
  University of Cambridge \\ 

  \And
  Stephen J. Price \\
  Department of Clinical Neurosciences \\
  University of Cambridge \\ 
  \And
  Carola-Bibiane Schönlieb \\
  Department of Applied Mathematics and Theoretical Physics \\
  University of Cambridge\\ 
}
\maketitle

\begin{abstract}
 We present an Expectation-Maximization (EM) Regularized Deep Learning (EMReDL) model for weakly supervised tumor segmentation. The proposed framework is tailored to glioblastoma, a type of malignant tumor characterized by its diffuse infiltration into the surrounding brain tissue, which poses significant challenge to treatment target and tumor burden estimation using conventional structural MRI. Although physiological MRI provides more specific information regarding tumor infiltration, the relatively low resolution hinders a precise full annotation. This has motivated us to develop a weakly supervised deep learning solution that exploits the partial labelled tumor regions. 

EMReDL contains two components: a physiological prior prediction model and EM-regularized segmentation model. The physiological prior prediction model exploits the physiological MRI by training a classifier to generate a physiological prior map. This map is passed to the segmentation model for regularization using the EM algorithm. We evaluated the model on a glioblastoma dataset with the pre-operative multiparametric and recurrence MRI available. EMReDL showed to effectively segment the infiltrated tumor from the partially labelled  region of potential infiltration. The segmented core tumor and infiltrated tumor demonstrated high consistency with the tumor burden labelled by experts. The performance comparisons showed that EMReDL achieved higher accuracy than published state-of-the-art models. On MR spectroscopy, the segmented region displayed more aggressive features than other partial labelled region. The proposed model can be generalized to other segmentation tasks that rely on partial labels, with the CNN architecture flexible in the framework.  

\end{abstract}

\section{Introduction}

Glioblastoma is the most common malignant primary brain tumor, characterized by poor outcomes \citep{Wen2020}. The first-line treatment includes maximal safe resection followed by chemoradiotherapy \citep{Stupp2005}, which requires an accurate tumor delineation to enhance the treatment efficacy and reduce the neurological deficits of patients \citep{Mazzara2004,Stupp2005}. As the manual delineation is often subjective and laborious, an automated tumor segmentation model is crucial in aiding clinical practice.
Currently, magnetic resonance imaging (MRI) is the mainstay for diagnosis, treatment planning, and disease monitoring of glioblastoma \citep{Weller2014,Weller2017,Wen2020} . It however remains a challenge to accurately segment the glioblastoma based on MRI \citep{Wadhwa2019}, mainly due to several reasons. Firstly, glioblastoma is characterized by diffuse infiltration into the surrounding brain, leading to a poorly demarcated tumor margin. Secondly, glioblastoma is highly heterogeneous with regard to the tumor location, morphology and intensity values. Thirdly, glioblastoma may demonstrate similar appearance with neurodegenerative or white matter pathologies. All of the above may pose significant challenges to a robust segmentation model.

Incorporating multiple MRI modalities is considered beneficial for tumor segmentation \citep{Ghaffari2020}. Clinically, the most commonly used sequences include T1-weighted, T2-weighted, post-contrast T1-weighted (T1C), and fluid attenuation inversion recovery (FLAIR) sequences. A multimodal brain tumor image segmentation (BraTS) challenge represents the collective efforts to develop segmentation models using a large glioblastoma dataset with multiple MRI sequences available \citep{Bakas2018}. A wide spectrum of models has since been proposed with dramatic success in performance \citep{Ghaffari2020}. Among these models, deep learning shows unique advantages in using multiple MRI sequences for tumor segmentation, compared to the traditional methods of using hand-crafted features. However, the BraTS dataset only includes the most widely used structural sequences, which was shown to be prone to the low specificity in targeting actual tumor infiltration \citep{Verburg2020}. Particularly, for the non-enhancing lesion beyond the contrast-enhancing margin, it remains challenging to differentiate the infiltrated tumor from edema, even combining all the structural sequences \citep{Verburg2020}. An effective imaging model with higher specificity in 
segmenting the infiltrated tumor is of crucial value for clinical decision making. 

An increasing amount of literature provides evidence that physiological MRI can facilitate the characterization of tumor infiltration \citep{Li2019,Yan2019}. In particular, diffusion and perfusion MRI can identify the infiltrated tumor beyond the contrast enhancement by offering parametric measures describing tumor physiology, which may complement the non-specificity of the structural sequences. Specifically, The diffusion MRI is the only imaging method of describing brain microstructure by measuring water molecule mobility \citep{jellison2004diffusion}, which can detect the subtle infiltration \citep{Li20192}, characterize tumor invasiveness \citep{Li20193} and predict tumor progression \citep{Yan2020}. On the other hand, as a widely used perfusion technique, dynamic susceptibility contrast (DSC) imaging can derive the relative cerebral blood volume (rCBV), mean transit time (MTT) and relative cerebral blood flow (rCBF), reflecting the aberrant tumor vascularization \citep{lupo2005dynamic}. Therefore, integrating physiological MRI into the tumor segmentation model shows potential to more accurately identify tumor infiltration. 

Here we proposed a deep learning model to automatically segment the core and infiltrated tumor based on both structural and physiological multiparametric MRI. We hypothesized that the physiological MRI information of the core tumor could be used to guide the deep learning model to segment the infiltrated tumor beyond the core tumor. In the next section, we summarize the related work of tumor segmentation, including both supervised and weakly supervised models.

\section{Related work}

Tumor segmentation is an active research field with a growing number of models proposed. These models can be generally classified into generative or discriminative models \citep{Ghaffari2020}. Typically, generative models rely on the prior knowledge of the voxel distributions of the brain tissue, which is derived from the probabilistic atlas \citep{Prastawa2004}, whereas the discriminative models rely on the extracted image features that could be mapped to the classification labels. In general, discriminative models show superior performance than generative models. Most successful discriminative approaches in the BraTS challenge \citep{Menze2015} are based on fully supervised convolutional neural networks (CNN).  

In BraTS 2014, a CNN-based model was firstly introduced. The top-ranked algorithm employed a 3D CNN model trained on small image patches, which consisted of four convolutional layers with six filters in the last layer corresponding to six labels \citep{Urban2014}. In BraTs 2015, a 2D CNN model with a cascaded architecture was proposed. Two parallel CNNs were employed to extract local and global features which were then concatenated and fed into a fully connected layer for classification \citep{Dutil2015}. In BraTS 2016, DeepMedic, a 3D CNN model of eleven layers with residual connections was proposed. Two pathways were employed to process the inputs in parallel, to increase the receptive field of the classification layer \citep{kamnitsas2016deepmedic}. In BraTS 2017, the Ensembles of Multiple Models and Architectures (EMMA) separately trained several models (DeepMedics, 3D FCN, and 3D U-net) using different optimization approaches, while the output was defined as the average to reduce bias from individual models \citep{kamnitsas2017ensembles}. The top-ranked model in BraTS 2018 proposed an asymmetric U-net architecture, where an additional variational auto-encoder branch was added to the shared encoder, providing additional regularization \citep{myronenko20183d,warrington2020xtract}. In BraTS 2019, the top-ranked model proposed a two-stage cascaded U-Net \citep{jiang2019two}. The first stage used a U-Net variant for preliminary prediction, whereas the second stage concatenated the preliminary prediction map with the original input images to refine the prediction. 

In summary, the above top-ranked models from the BraTS depict the advantages of CNN-based segmentation model, which highlights the capacity of feature extraction of CNN. Further, to enhance the model performance or reduce the computational cost, various techniques were employed to improve the backbone CNN by a series of procedures, e.g., increasing network depth or width, optimizing the loss function, increasing receptive fields, or adopting an ensemble model. For more details of the BraTS models, please refer to \citep{Bakas2018,Ghaffari2020}.
All these state-of-the-art models heavily rely on the full classification labels to train a model that could approximate the accuracy of experts. The infiltrative nature of glioblastoma, however, poses significant challenges to accurate delineation of the interface between tumor and healthy tissue. Although the binary contrast-enhancement provided a reference for “core tumor”, the surrounding non-enhancing region, regarded as the edema in BraTS labels, has established as diffusively infiltrated with tumor. 

As outlined in the previous section, multiparametric MRI allows more accurate identification of the non-enhancing  infiltrated tumor. Nevertheless, the low resolution of physiological MRI hinders the precise annotation based on these images. A full annotation based on physiological MRI therefore is prone to the subjective errors, even by experienced clinical experts. As a result, those models with high reliance on the full labels may not be suitable for segmented the infiltrated tumor.  

Other studies investigated the feasibility of delineating tumor infiltration based on the weak labels of cancerous and healthy tissues. \citep{Akbari2016} proposed a tumor infiltration inference model using the physiological and structural MRI \citep{Akbari2016}. Two types of weak labels were used, i.e., one scribble immediately adjacent to the enhancing tumor and another scribble near the distal margin of the edema. These two scribble regions, representing the tissue near and far from the core tumor respectively, were hypothesized to correspondingly have higher and lower tumor infiltration. The classifier was trained based on the weak labels using the support vector machine (SVM) on a voxel-by-voxel basis which yielded a infiltration probability map. The model achieved excellent performance and was subsequently validated by another cohort and the tumor recurrence on the follow-up scans. 

Although in relatively small sample size, this study underpinned the advantage of physiological MRI in identifying tumor infiltration and supported the feasibility of weakly supervised learning models to tackle the challenge of lacking precise full annotations. The proposed model, however, ignored the spatial continuity of tumor infiltration. The CNN model could empower the weakly supervised learning model \citep{Chan2020} by effectively extracting multiparametric MRI features with spatial information.

However, directly training a weakly supervised CNN model using a partial cross-entropy (PCE) loss may lead to poor boundary localization of saliency maps \citep{zhang2020weakly}. To mitigate this limitation, additional regularization is often employed. For instance, \citep{tang2018normalized} introduced a normalized cut loss as a regularizer with a partial cross-entropy loss. \citep{Kervadec2019} introduced a regularization term constraining the size of the target region that was combined with a partial cross-entropy loss. \citep{Roth2019} used the random walker algorithm to generate the pseudo full label from the partial labels and then constructed the regularized loss by enforcing the CNN outputs to match the pseudo labels. The results of above studies supported the usefulness of additional regularizers in the weakly supervised models. Due to the advantages of physiological MRI in detecting tumor infiltration, here we hypothesized that a regularizer from the physiological MRI could enhance the weakly supervised model for segmenting the infiltrated tumor by incorporating domain-specific information. 

We sought to propose a CNN-based weakly supervised model, in which a regularization term was constructed by incorporating the prior information obtained from the physiological MRI by an prediction model through an expectation-maximization (EM) framework. We evaluated the model validation using tumor recurrence on follow-up scans and MR spectroscopy that non-invasively measures the metabolic alternation.   
The remainder of this paper is organized as follows: Section 3 will describe the overall study design, main components of the proposed framework and the performance evaluation of the model. Section 4 gives details of the dataset and the implementation of the experiments. Section 5 will provide the results and discussion followed by the conclusions in Section 6.

\section{Methodology}
Consider multi-parametric MRI from $N$ (patients) training samples $\mbX = \{X_1, X_2, \cdots, X_N \}$, including both structural sequences (T1-weighted, T2-weighted, T1C and FLAIR) and physiological sequences (diffusion and perfusion MRI), denoted as $\mbX_{s}$ and $\mbX_{p}$, respectively. From a clinical perspective, three regions of interest (ROI) can be delineated (Figure 1): 
\begin{list}{$\circ$}{} 
\item 
ROI1: core tumor, which is the contrast-enhancing tumor region on T1C images and the surgery target for clinical practice; 
\item
ROI2: potential infiltrated region, which is hyperintensities in FLAIR images outside of ROI1. We are  specifically interested in this region as it represents the clinically extendable treatment target; 
\item
ROI3: normal-appearing region on both T1C and FLAIR sequences. 
\end{list}

All preoperative MRI sequences have been co-registered using affine transformation, and the The labels for voxels in ROI1 and ROI3 are observable, denoted as $\mby_o$. A voxel label $\mby_o$ is a value either $1$ or $0$, and $\mby_o=1$ indicates confirmed tumor voxel (ROI1) and $\mby_o=0$ represents a voxel from the normal appearing brain region (ROI3). The infiltration percentage in ROI2 is a latent term and $\mby_u$ indicates the voxel-wise labels for ROI2, for which the probability distribution $P\left(\mby_u=1\right)$ need to be estimated. 

\graphicspath{ {./images/} }
\begin{figure}[!h]
\centering
\includegraphics[scale=.36]{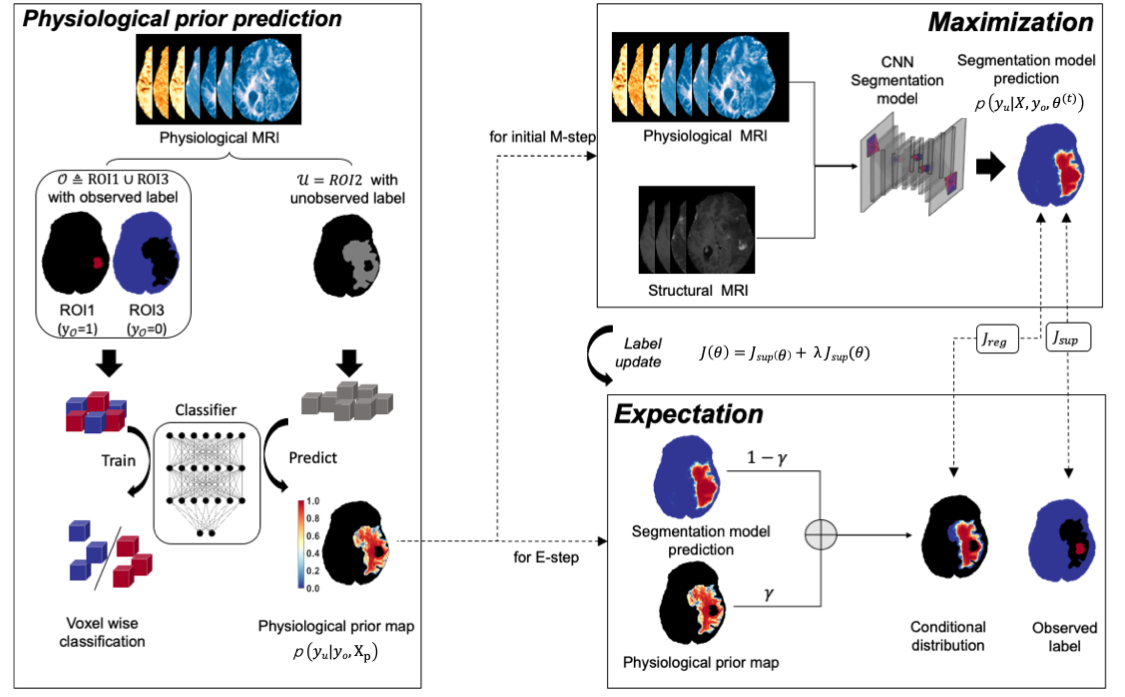}
\caption{Diagram of the proposed method. The left panel describes the physiological prior prediction process. A classifier is trained to generate physiological prior map. The right panel depicts the EM regularized CNN model training process. The Expectation-Maximization (EM) framework is used to fulfill and optimize the weakly supervised model, where a CNN model is trained in the M-step and the distribution of unobserved ROI2 are estimated in the E-step. $J(\mtheta)$ is the loss term of CNN model, and is calculated using scaled summation of both $J_{\text{reg}}(\mtheta)$ and $J_{\text{sup}}(\mtheta)$. The term $J_{\text{reg}}(\mtheta)$ denotes the regularized loss generated by the conditional distribution computed by Equation \eqref{Eq:EMLossReg}, and $J_{\text{sup}}(\mtheta)$ denotes the supervised loss from the observed labels $\mby_o$.}
\label{Fig:ill}
\end{figure}

\subsection{Overview of EMReDL method}

Our goal is to segment the core and infiltrated tumor using the model trained by the existing MRI data $\mbX$ and its corresponding observed labels $\mby_o$. For the standard supervised CNN models, full training labels are necessary to be used as the `ground-truth' to train the weights $\mtheta$ of the CNN. In our proposed application, however, as it is not possible to obtain a full annotation for the unobserved labels $\mby_u$, which renders a supervised CNN training inappropriate. In this paper, we cast the underlying problem into a weakly supervised learning problem by leveraging the EM algorithm, which can recursively estimate both the CNN parameters (M-step) and the unobserved labels (E-step) in the proposed segmentation problem. The problem can now be treated as training a CNN segmentation model using partial labels.

As shown in Figure \ref{Fig:ill}, the proposed method consists of two main components: physiological prior prediction model (left panel) and EM-regularized segmentation model (right panel). The left panel takes in physiological MRI information to train a classifier and generate initial voxel-wise estimate of the unobserved labels in ROI2. The estimated label information is then passed into the right panel to improve the prediction performance of the segmentation model. Specifically, the label information is used to initialize ROI2 labels in the CNN model training in M-step, and is also integrated into E-step to recursively update the estimation of the unobserved label $\mby_u$. The expected outcome of the right panel is a trained CNN segmentation model that can effectively distinguish the infiltrated tumor from the non-cancerous edema. 

The pipeline introduced in Figure \ref{Fig:ill} can be further generalized to other similar segmentation problems with partially unobserved labels. Both the classifier in the left panel and the CNN segmentation model in the right panel are flexible to be replaced by other feed-forward deep learning models or CNN models with architectures other than the ones used in this paper. Given this, we will not explicitly describe detailed architecture of the CNN models used in the proposed method. 

\subsection{Voxel-wise physiological prior map}
Specific information contained in the physiological MRI can be used to identify potential tumor infiltration. However, its image resolution is often lower than that of the structural MRI. In this work, a voxel-wise \textit{physiological prior map} is derived to depict the prior distribution of the unknown voxel labels $\mby_u$ in ROI2 by leveraging the information extracted from the physiological MRI data. Specifically, the voxel-wise prior map is approximated by a simple binary neural network (NN) classifier trained on both the physiological MRI $\mbX_p$ and the observed labels $\mby_o$. The derived physiological prior map is defined as $p(\mby_u | \mbX_p, \mby_o)$, which then serves as prior information in the later EM optimization step for weakly supervised image segmentation. 

One major drawback of this voxel-wise binary classification model is its negligence of prediction reliability from the trained deterministic classifier. The classifier `must' assign a binary label to each voxel even with low confidence, and this will certainly bring extra bias to the EM optimization step. To mitigate this issue, we modify the binary NN classifier to estimate both the binary prior map and its corresponding uncertainty $\delta(\mby_u | \mbX_{p}, \mby_o)$. This is achieved by a well established evidential deep learning (EDL) algorithm \citep{Sensoy2018edl}, which modifies the original architecture of the NN rather than constructing a separate complex model from scratch. It simply replaces the output layer of NN by a ReLU activation function, and modifies the corresponding loss function by a summation of expected mean squared error and a Kullback–Leibler (KL) divergence term based on the output of ReLU function. For instance, for a K-class classification task with a standard Softmax function, the Softmax outputs are replaced by ReLU outputs and then used to model an order K Dirichlet distribution. The corresponding uncertainty of the predictions can finally be computed based on the Dirichlet model. 

The derivation of both the physiological prior map and uncertainty is described in Algorithm \ref{alg1:PriorPred}, where $t$ represents the number of iterations. The original binary classifier is constructed by a fully connected feed-forward NN with two hidden layers. The number of hidden neurons is set equal to the number of input features (physiological MR signatures), and the ReLU activation is adopted across all layers. The loss function is modified mainly following the original EDL steps (see \citep{Sensoy2018edl} for details.) with value adaption/normalisation to this specific problem. Since the proposed NN classifier only makes binary prediction (i.e., $K=2$), the Dirichlet distribution in a multi-class EDL classifier turns to its 2 dimensional form of Beta distribution after the EDL modification.

\begin{center}
\begin{minipage}{.8\linewidth}
\begin{algorithm}[H]
\caption{Derivation of the physiological prior map}
\label{alg1:PriorPred}
\tcp{Training of EDL modified binary classifier}
\KwInput{MRI samples $\mbX_p$, class labels $\mby_o$}
Randomly initialize neuron weight set $\momega$ at $t=0$\\
\For{$t = 1,2, \cdots$ }{
Compute predictions $\hat{\mby}_o$ according to $\momega^{(t-1)}$ \\
Compute error between $\hat{\mby}_o$ and $\mby_o$ \\
Update $\momega^{(t)} \leftarrow \momega^{(t-1)} $}
\tcp{Repeat prediction}
\ForAll{Voxels in ROI2}{
Predict $\hat{\mby}_u$ and estimate associated uncertainty score \\
}
Derive prior map $p(\mby_u | \mbX_p, \mby_o)$ \\
Derive uncertainty distribution $\delta(\mby_u | \mbX_{p}, \mby_o)$ 
\end{algorithm}
\end{minipage}
\end{center}

\subsection{Segmentation using EM-regularized weakly supervised learning}
In this component, a segmentation model constructed by a standard U-Net CNN architecture is trained for tumor segmentation. Different from the physiological prior prediction model, the segmentation model is trained using both physiological MRI $\mbX_p$ and structural MRI $\mbX_s$. An EM algorithm is employed in this step to recursively perform mutual optimization on both the unobserved label $\mby_u$ and CNN model weights. We firstly define the likelihood function as:
\begin{align}
    L(\mtheta) = p(\mby_u, \mby_o | \mbX, \mtheta),
    \label{Eq: EMLikelihood}
\end{align}
for which the maximum likelihood estimate with respect to the weights $\theta$ (of CNN) can be computed by integrating out the unknown term $\mby_u$ and maximizing the marginal distribution:
\begin{align}
    p(\mby_o | \mbX, \theta) = \int p(\mby_u, \mby_o | \mbX, \mtheta) d \mby_u.
\end{align}
Nevertheless, the integral is often intractable and exact integration over all possible $\mby_u$ values is challenging. EM algorithm maximize the likelihood $L(\mtheta)$ by iteratively estimating the unknown term $\mby_u$ in the expectation step (\textit{E-step}) and $\mtheta$ in the maximization step (\textit{M-step}). See \citep{mclachlan2007algorithm} for details of the standard EM algorithm. 

In this paper, EM performs E-step by defining 
\begin{align}
    Q(\mtheta | \mtheta^{(t)}) = \mathE_{\mby_u | \mby_o, \mbX, \mtheta^{(t)}} [\log L(\mtheta)]
    = \int \log p\left(\mby_u, \mby_o | \mbX, \mtheta\right)d p(\mby_u | \mby_o, \mbX, \mtheta^{(t)}), \label{Eq:EMEstep2} 
\end{align}
where $\mtheta^{(t)}$ denotes the estimated CNN weights in iteration $t$. $Q(\mtheta | \mtheta^{(t)})$ computes the expectation of the log-likelihood of function $L(\mtheta)$ with respect to the conditional distribution $p(\mby_u | \mby_o, \mbX, \mtheta^{(t)})$, which is defined as:
\begin{align}
    p(\mby_u | \mby_o, \mbX, \mtheta^{(t)}) = (1 - \gamma_t) p(\mby_u | \mbX_{p}, \mby_o) + \gamma_t p(\mby_u | \mbX, \mtheta^{(t)}).
    \label{Eq:EMUpdate}
\end{align}
The former term on the RHS is the physiological prior map generated by the binary classifier and the latter term is the predicted probabilistic labels in the $t$th iteration of segmentation model. $\gamma_t$ denotes a voxel-wise binary coefficient, which will be used to integrate the physiological prior map and the prediction of segmentation model. $\gamma_t$ is a binary indicator function defined as follows:
\begin{align}
    \gamma_t = I_A(\mby_u),
    \text{where } A=\{g(p(\mby_u | \mbX_{p}, \mby_o), p(\mby_u | \mbX, \mtheta^{(t)})) \leq \delta(\mby_u | \mbX_{p}, \mby_o)\}.
    \label{Eq:Indicator}
\end{align}
$g(\cdot)$ is a predefined function measures the dissimilarity between the prediction of segmentation model and the prior map. The predicted labels from segmentation model is accepted as the pseudo training label for next epoch only if the dissimilarity value is smaller or equal to the uncertainty score defined by $\delta(\mby_u | \mbX_{p}, \mby_o)$.

M-step is to maximize the above quantity to derive new estimate $\mtheta^{(t+1)}$:
\begin{align}
    \mtheta^{(t+1)} = \arg \max_{\mtheta} Q(\mtheta | \mtheta^{(t)}).
    \label{Eq:EMMstep}
\end{align}

From the perspective of loss function in CNN model training, Equation \eqref{Eq:EMMstep} can also be reformulated as a regularization term $J_{\text{reg}}(\mtheta) \triangleq - Q(\mtheta | \mtheta^{(t)})$, which helps to mitigate potential poor boundary localization issue in CNN training. The overall loss function $J(\mtheta)$ is can then be defined as  
\begin{align}
    J(\mtheta) = J_{\text{sup}}(\mtheta) + \lambda J_{\text{reg}}(\mtheta), 
    \label{Eq:EMReDLLoss}
\end{align}
where $\lambda \in [0,1]$ is the pre-defined regularization coefficient. $J_{\text{sup}}(\mtheta) = -\log\left(p(\mby_o | \mbX, \mtheta)\right)$ is the standard negative log-likelihood loss. In fact, $J_{\text{reg}}(\mtheta)$ penalizes the prediction error in ROI2 and $J_{\text{sup}}(\mtheta)$ penalizes the prediction error in ROI1 and ROI3. 

Now Equation \eqref{Eq:EMEstep2} can be rewritten as:
\begin{align}
    Q(\mtheta | \mtheta^{(t)}) 
    = \int \left(\log    p\left(\mby_u| \mbX, \mtheta\right) + \log p\left(\mby_o |\mby_u, \mbX \right)    \right) dp(\mby_u | \mby_o, \mbX, \mtheta^{(t)}).
    \label{Eq:EMReg1} 
\end{align}
$J_{\text{reg}}(\mtheta)$ can be further simplified with respect to $\mtheta$:
\begin{align}
    J_{\text{reg}}(\mtheta) = - \int \log    p\left(\mby_u| \mbX, \mtheta\right)  dp(\mby_u | \mby_o, \mbX, \mtheta^{(t)}) + C,
    \label{Eq:EMLossReg}
\end{align}
where $C$ is a constant. $J_{\text{reg}}(\mtheta)$ corresponds to cross-entropy loss of the predictive distribution $p\left(\mby_u| \mbX, \mtheta\right)$ for ROI2, while $J_{\text{sup}}(\mtheta)$ can also be treated as partial cross-entropy loss for ROI1 and ROI3. The step-by-step procedure of the proposed EM-Regularized weakly-supervised segmentation is shown in Algorithm \ref{alg2:EMReDL}.

\begin{center}
\begin{minipage}{.8\linewidth}
\begin{algorithm}[H]
\caption{EM-Regularized weakly-supervised segmentation}
\label{alg2:EMReDL}
\KwInput{MRI samples $\mbX = \{\mbX_{s}, \mbX_{p}\}$, class labels $\mby_o$}
Initialize weight $\mtheta$, coefficient $\lambda$, and indicator $\gamma$ at $t=0$\\
Compute $p(\mby_u | \mbX_p, \mby_o)$ from Algorithm \ref{alg1:PriorPred} \\
\Repeat{Convergence reached, obtain optimal $\mtheta^{*}$}{
\tcp{E-step}
Determine $\gamma_t$ by Eq~\eqref{Eq:Indicator} \\
Compute $p(\mby_u | \mby_o, \mbX, \mtheta^{(t-1)}) $ by Eq~\eqref{Eq:EMUpdate} \\
\tcp{M-step}
Compute $J_{\text{reg}}(\mtheta^{t})$ by Eq~\eqref{Eq:EMLossReg} \\
Obtain $J(\mtheta^{t-1}) = J_{\text{sup}}(\mtheta^{t-1}) + \lambda J_{\text{reg}}(\mtheta^{t-1})$ by Eq~\eqref{Eq:EMReDLLoss} \\
Update $\mtheta^{(t)} \leftarrow \mtheta^{(t-1)}$ according to Eq~\eqref{Eq:EMMstep} \\
}
Extract infiltration region $\widehat{R}_{seg}$ by thresholding $p(\mby|\mbX,\mtheta^{*})$ \\
\KwRet{$p(\mby|\mbX,\mtheta^{*})$, $\widehat{R}_{seg}$}
\end{algorithm}
\end{minipage}
\end{center}

\subsection{Model evaluation}

We validated the proposed model using tumor burden, tumor recurrence and magnetic resonance spectroscopy (MRS). To examine the usefulness of the regularizer, we compared our model performance with the baseline model which employed the U-net with a partial cross-entropy (PCE) loss without the additional regularizer from the physiological prior. We also compared our model with a traditional infiltration prediction model with multiparametric MRI \citep{Akbari2016} and other state-of-the-art weakly supervised segmentation (WSS) methods trained with partial pixel-level labels, including \cite{tang2018normalized,Kervadec2019,Roth2019}.

\textit{1) Tumor burden estimation}

Tumor burden is crucial for patient risk stratification and treatment planning. The volume of finally segmented tumor can be divided into core tumor burden (the volume of delineated tumor in ROI1) and infiltrated tumor burden (the volume of delineated tumor in ROI2). A linear regression was used to test the consistency of the segmented volumes from different models with the ground truth. For the core tumor (ROI1), the volume of the manual label was used as ground truth. For the infiltrated tumor, the volumme of the recurrence within the potential infiltrated region (ROI2) was considered as the ground truth.

\textit{2) Recurrence prediction}

The finally segmented tumor region was examined in the prediction of whole tumor region and tumor recurrence region in the follow-up MRI of 68 patients. The potential infiltrated region (ROI2) on the pre-operative images was  divided into recurrence region $R_{recur}$ and non-recurrence region $R_{recur}^{C}$, according to the manual label, where ${C}$ represents the complementary operation.

For each patient with follow-up MRI, the pre-operative contrast-enhancing core tumor (ROI1) on T1C image was denoted as $R_{CE}$, therefore the whole tumor region was defined as 
$R_{whole}$ = $R_{recur}$ $\cup$ $R_{CE}$. The segmented tumor area {$\widehat{R}_{seg}$} and normal-appearing area $\widehat{R}_{seg}^{C}$ can be derived automatically by thresholding the tumor infiltration probability map that was produced by EMReDL. Finally, The sensitivity and specificity of predicting whole tumor region were defined as:

\begin{equation}
{\rm Sensitivity}= \frac{ {\rm Volume}\left(R_{whole}\cap\widehat{R}_{seg}\right)}{ {\rm Volume} \left(R_{total}\right)}
\label{Eq:Sensi}
\end{equation}
\begin{equation}
{\rm Specificity}= \frac{ {\rm Volume} \left(R_{recur}^{C}\cap\widehat{R}_{seg}^{C}\right)}{ {\rm Volume} \left(R_{recur}^{C}\right)} 
\end{equation}

After calculating the sensitive and specificity, the optimum threshold $\tau$ for discriminating predicted infiltration mask was chosen by maximizing the Youden Index of the ROC curves. When evaluating the sensitivity and specificity for predicting tumor recurrence region, we replaced the $R_{whole}$ with $R_{recur}$ in \eqref{Eq:Sensi}.

\textit{3) Magnetic resonance spectroscopy validation}

The metabolic signature was compared for the infiltrated region and non-infiltrated region segmented by the algorithm in the potential infiltrated region (ROI2). The metabolic measures, including Choline (Cho), N-acetylaspartate (NAA) and Cho/NAA were calculated for the infiltrated region and non-infiltrated region, respectively. To account for the resolution difference between T2 and MRS space, all co-registered data were projected to MRS space according to their coordinates. The proportion of T2-space tumor pixels occupying each MRS voxel was calculated. Paired t-test was used to compare the metabolic measures of the infiltration and non-infiltration regions.

\section{Experiments}
\label{others}

\subsection{Data description}

This study was approved by the local institutional review board and informed consent was obtained from all patients. A total of 115 glioblastoma patients was prospectively recruited for maximal safe resection. Each patient underwent pre-operative multiparametric MRI, using a 3-Tesla MRI system (Magnetron Trio; Siemens Healthcare, Erlangen, Germany) with a standard 12-channel receive-head coil. The sequences included T1, T1C, T2, T2-FLAIR, diffusion imaging, Dynamic Susceptibility contrast-enhanced imaging and multivoxel 2D 1H-MRS chemical shift imaging. Patient was treated and followed up by the multidisciplinary team (MDT) according to the clinical guidelines. The extent of resection was assessed according to post-operative MRI within 72 hours. During the follow up of patients, clinical and radiological data were incorporated according to the Response Assessment in Neuro-oncology criteria. For recurrence analysis, the follow-up MRI data included 68 patients who received the complete resection, which is defined clinically as a complete resection of contrast-enhancing tumor (ROI1).

\subsection{Image pre-processing}
\textit{1) Multiparametric MRI processing }

Diffusion MRI was processed using the diffusion toolbox (FDT) in FSL v5.0.8 (FMRIB Software Library, Centre for Functional MRI of the Brain, Oxford, UK). After normalization and eddy current correction, parametric maps of fractional anisotropy (FA), mean diffusivity (MD), p (isotropy) and q (anisotropy) were calculated as previously described \citep{li2019characterizing, li2019intratumoral}. DSC was processed using the NordicICE (NordicNeuroLab, Bergen, Norway), with arterial input function automatically defined and leakage corrected. The parametric maps of rCBV, MTT and rCBF maps were calculated. The MRS data were processed using LCModel (Provencher, Oakville, Ontario). All metabolites were calculated as a ratio to creatine (Cr).

\textit{2) Image co-registration}

All pre-operative parametric maps were co-registered to the T2 space using FSL linear image registration tool (FLIRT) with an affine transformation. For the co-registration of the recurrence image (follow-up MR scan) to the pre-operative images, the recurrence T1C images were non-linearly co-registered to the pre-operative T2 images using the Advanced Normalization Tools (ANTs), with the pre-operative lesion masked out.

\textit{3) Image normalization}

All MRI from different patients were normalized using the histogram matching method. Specifically, for each sequence, the image histograms for all patients were calculated, where the histogram closest to the averaged histogram was determined as the reference and normalized to [0, 1]. Finally, other image were matched to the reference histogram using histogram matching implemented in MATLAB.
\subsection{Labelling of pre-operative and recurrence tumor}
Preoperative tumor and recurrence regions were manually delineated on the T1C and FLAIR images using the 3D slicer v4.6.2 (https://www.slicer.org/). The delineation was independently performed by a neurosurgeon (XX) and reviewed by a neuroradiologist (XX). Each rater used consistent criteria in each patient and was blinded to patient outcomes. The contrast-enhancing core tumor (CE ROI, ROI1) was defined as the regions within the contrast-enhancing margin on T1C images. The FLAIR ROI was defined as the hyperintensities on FLAIR images. Finally, the peritumoral ROI (ROI2) were defined as the FLAIR ROI outside of contrast-enhancing regions, obtained by a Boolean subtraction of FLAIR and CE ROIs.

\subsection{Implementation details of EMReDL}
We divided the complete dataset into two sets randomly: ~50\% as the training set from 57 patients (35 patients with recurrence images) and ~50\% as the testing set from 58 patients (33 patients with recurrence images). Each patient has 23 slices of multiparametric MR images along the axial axis for both preoperative and follow-up scan. For the training set, 75\% of the data was randomly chosen and used for model training and the remaining 25\% was used for model validation.

For the training of physiological prior prediction model, the multiparametric MRI feature vectors for all voxels in the ROI1 and ROI3 were used as the input of the fully connected network. Adam optimizer was applied to train the model with initial learning rate set to $10^{-4}$, and the model was trained for 1000 epochs using mini-batches of size $5\times10^{4}$. To tackle the class imbalance problem, equal numbers of majority- and minority-class samples were randomly selected for each mini-batch. Finally, the model with smallest validation error was adopted.

After the training of the physiological prior prediction model, a physiological prior map with the tumor infiltration probability was obtained for ROI2. The EM-regularized weakly supervised segmentation model was trained to minimize the loss function \eqref{Eq:EMReDLLoss} for 200 epochs using Adam optimizer with initial learning rate of $10^{-4}$, and mini-batch size of 8. The parameter $\lambda$ in loss function \eqref{Eq:EMReDLLoss} is set to 1 in our experiment. For the training of the first epoch, $\gamma_1$ in Equation \eqref{Eq:EMUpdate} is set 0, which means the prior infiltration probability was used as the probabilistic training labels in ROI2, i.e., the potential infiltration regions. Afterwards, the probabilistic training labels were updated for each epoch based on Equation \eqref{Eq:EMUpdate}. The model with lowest validation error was finally chosen. The proposed method was implemented using TensorFlow Library. 


\section{Results and Discussion}
The experiment results showed that the proposed weakly supervised model achieved high accuracy in segmenting the core and infiltrated tumor area, which can be validated by the tumor burden estimation, tumor recurrence prediction and identification of invasive areas in MRS. The results are presented in below. 

\subsection{Tumor burden estimation}
The volume of the segmented regions were calculated for different models as tumor burden (Table \ref{tab:Table4}). For the core tumor, the results showed that all CNN models achieved comparable volumes with the grund truth, highlighting the capacity of CNN in core tumor segmentation. For the infiltrated tumor, our results showed EMReDL achieved most similar results with the recurrence volume.

\begin{table}[!h]
{\centering
\caption{Tumor burden estimation of different models}
\label{tab:Table4}
\resizebox{\textwidth}{!}{%
\begin{tabular}{ccccccccc}
\hline
\multicolumn{2}{c}{} & \begin{tabular}[c]{@{}c@{}}Ground \\ truth\end{tabular} & Baseline & \begin{tabular}[c]{@{}c@{}}Comparison \\ model 1\end{tabular} & \begin{tabular}[c]{@{}c@{}}Comparison \\ model 2\end{tabular} & \begin{tabular}[c]{@{}c@{}}Comparison \\ model 3\end{tabular} & \begin{tabular}[c]{@{}c@{}}Comparison \\ model 4\end{tabular} & EMReDL \\ \hline
\multirow{2}{*}{\begin{tabular}[c]{@{}c@{}}Core \\ tumor\end{tabular}} & Training & 45.4±29.4 & 44.8±29.2 & 33.7±18.9 & 45.0±29.2 & 43.6±27.7 & 43.4±27.7 & 44.3±28.7 \\
 & Testing & 48.8±29.7 & 46.8±29.0 & 36.4±20.5 & 46.7±28.8 & 45.7±27.7 & 43.8±26 & 45.0±27.1 \\ \hline
\multirow{2}{*}{\begin{tabular}[c]{@{}c@{}}Infiltrated \\ tumor\end{tabular}} & Training & 17.9±16.2 & 9.4±6.2 & 31.4±22.1 & 9.1±5.2 & 20.9±10.5 & 16.0±10.8 & 17.5±17.5 \\
 & Testing & 24.0±19.3 & 13.2±18 & 34.8±26.5 & 12.5±18.4 & 22.4±15.1 & 20.2±18.5 & 24.2±22.4 \\ \hline
\end{tabular}%
}
}
{Unit: $cm^3$; Baseline: U-net trained with PCE; Comparison model 1: SVM classifier based recurrence prediction \citep{Akbari2016}; Comparison model 2: Normalized cut loss based WSS \citep{tang2018normalized}; Comparison model 3: Size-constrained loss based WSS \citep{Kervadec2019}; Comparison model 4: Random walker regularization based WSS \citep{Roth2019}.}
\end{table}

We also performed the linear regression analysis between the tumor burden estimated from the models with the ground truth (Table \ref{tab:CorrTumorBurden}). The results showed that for the core tumor, all tested models showed consistency in core tumor burden estimation. However, for the infiltrated tumor, EMReDL achieved considerable improvements over other tested models. 

\begin{table}[!h]
{\centering
\caption{Linear correlations of tumor burden from ground truth and segmentation}
\label{tab:CorrTumorBurden}
\resizebox{\textwidth}{!}{%
\begin{tabular}{cccccccc}
\hline
\multicolumn{2}{c}{} & Baseline & \begin{tabular}[c]{@{}c@{}}Comparison \\ model 1\end{tabular} & \begin{tabular}[c]{@{}c@{}}Comparison \\ model 2\end{tabular} & \begin{tabular}[c]{@{}c@{}}Comparison \\ model 3\end{tabular} & \begin{tabular}[c]{@{}c@{}}Comparison \\ model 4\end{tabular} & EMReDL \\ \hline
\multirow{2}{*}{\begin{tabular}[c]{@{}c@{}}Core\\ tumor\end{tabular}} & $R$ & 0.998 & 0.939 & 0.998 & 0.989 & 0.990 & 0.995 \\
 & $p$ & <0.001 & <0.001 & <0.001 & <0.001 & <0.001 & <0.001 \\ \hline
\multirow{2}{*}{\begin{tabular}[c]{@{}c@{}}Infiltrated\\ tumor\end{tabular}} & $R$ & 0.641 & 0.839 & 0.549 & 0.842 & 0.859 & 0.978 \\
 & $p$ & <0.001 & <0.001 & <0.001 & <0.001 & <0.001 & <0.001 \\ \hline
\end{tabular}%
}}
{$R$: Pearson correlation coefficient; $p$: $p$-value (F-test for zero slope); Baseline: U-net trained with PCE; Comparison model 1: SVM classifier based recurrence prediction; Comparison model 2: Normalized cut loss based WSS; Comparison model 3: Size-constrained loss based WSS; Comparison model 4: Random walker regularization based WSS.}
\end{table}

\subsection{Recurrence prediction}
Firstly, we compared the performance of the baseline model and EMReDL. The quantitative comparison results of the EMReDL and baseline model are in Table \ref{tab:Table1}. The ablation experiment showed that EMReDL achieved higher accuracy in predicting tumor recurrence compared to the baseline model which employed the U-net architecture with PCE loss. The results suggest the usefulness of incorporating the additional regularizer constructed from the physiological MRI. Of note, the baseline model achieved slightly higher specificity in segmenting whole tumor, but much lower sensitivity than EMReDL, which is mainly due to the smaller segmented region by baseline model.

\begin{table}[!h]
\centering
\caption{Comparisons of baseline and EMReDL in predicting recurrence and whole tumor region}
\label{tab:Table1}
\begin{tabular}{cccccc}
\hline
\multicolumn{2}{c}{} & \multicolumn{2}{c}{Whole tumor region} & \multicolumn{2}{c}{Recurrence region} \\ \hline
\multicolumn{2}{c}{} & Baseline & EMReDL & Baseline & EMReDL \\ \hline
\multirow{2}{*}{AUC} & Train & 0.897 & 0.971 & 0.674 & 0.915\\
 & Test & 0.890 & 0.965 & 0.707 & 0.938\\ \hline
\multirow{2}{*}{Sensitivity} & Train & 0.789 & 0.906 & 0.463 & 0.809 \\
 & Test & 0.772 & 0.898 & 0.523 & 0.876\\ \hline
\multirow{2}{*}{Specificity} & Train & 0.929 & 0.918 & 0.868 & 0.890 \\
 & Test & 0.926 & 0.916 & 0.866 & 0.889\\ \hline
\multirow{2}{*}{\begin{tabular}[c]{@{}c@{}}Youden \\ index\end{tabular}} & Train & 0.697 & 0.825 & 0.331 & 0.699\\
 & Test & 0.718 & 0.813 & 0.389 & 0.765\\ \hline
\multirow{2}{*}{MCC} & Train & 0.716 & 0.823 & 0.353 & 0.677\\
 & Test & 0.689 & 0.808 & 0.414 & 0.746\\ \hline
\multirow{2}{*}{Dice} & Train & 0.745 & 0.849 & 0.339 & 0.621\\
 & Test & 0.733 & 0.846 & 0.408 & 0.711\\ \hline
\end{tabular}

{AUC: area under the curve. MCC: Matthews correlation coefficient}
\end{table}

\begin{figure}[!h]
\centering
 \includegraphics[scale=.29]{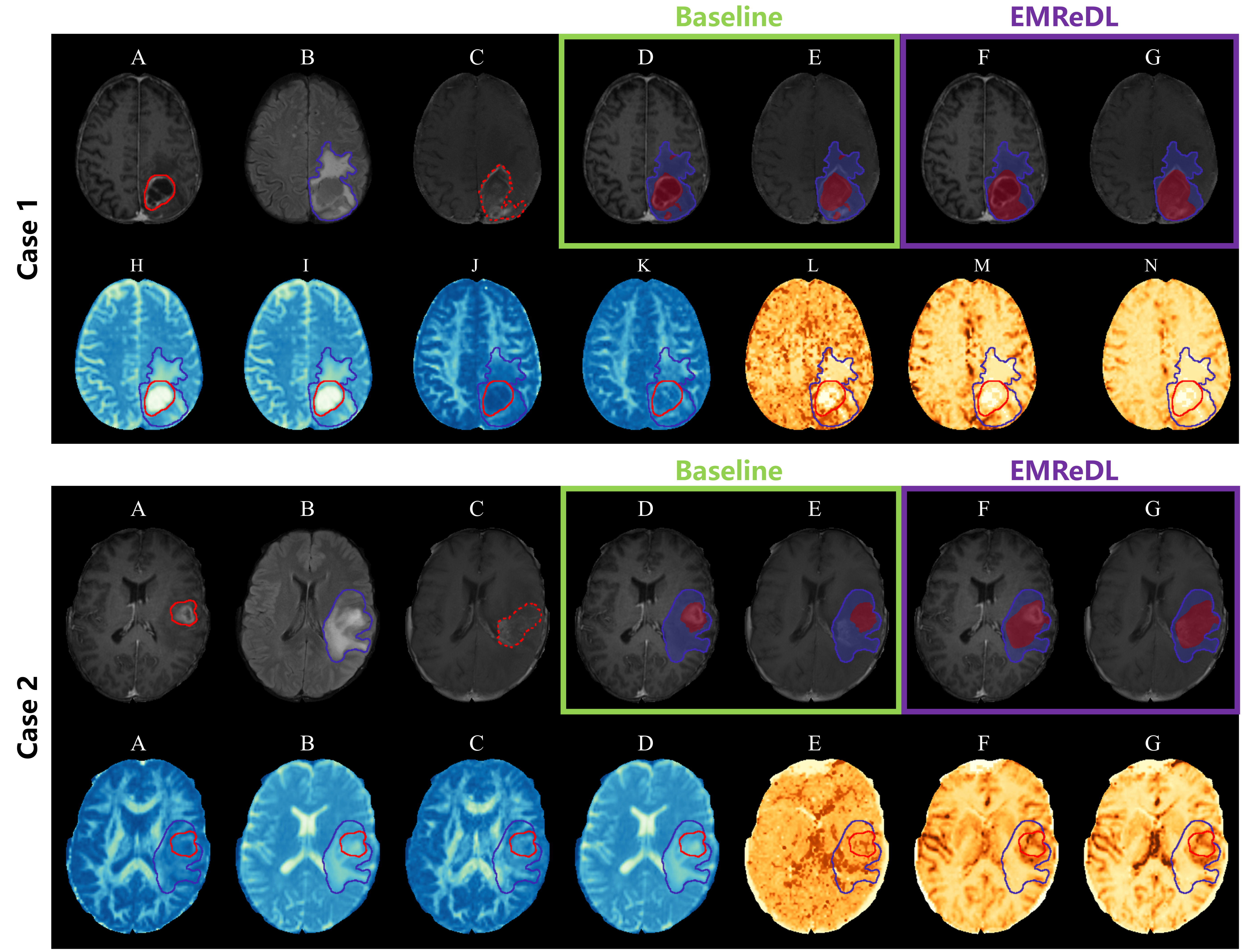}
\caption{ Typical segmentation results of baseline model and EMReDL for two cases. For both cases, A: preoperative T1C (red line: contrast-enhancing core tumor), B: preoperative FLAIR (blue line: peritumoral non-enhancing region), C: recurrence T1C (red dotted line: recurrence region); D$\sim$E: Segmentation result (red region) of baseline model and peritumoral non-enhancing region (blue region) overlaid on both preoperative and recurrence T1C images. F$\sim$G: Segmentation result (red region) of EMReDL and peritumoral non-enhancing region (blue region) overlaid on both preoperative and recurrence T1C images. H-N: preopearive FA, MD, DTI-q, DTI-p, MTT,  rCBF and rCBV images in sequence.}
\label{Fig:Figure2}
\end{figure}

Figure \ref{Fig:Figure2} presents two typical examples of infiltration area predicted by the EMReDL and the baseline model. The boundaries of core tumor (red line), peritumoral non-enhancing region (blue line), and recurrence region (red dotted line) are overlaid on preoperative T1C (Figure \ref{Fig:Figure2}A), FLAIR (Figure \ref{Fig:Figure2}B)  and recurrence T1C image (Figure \ref{Fig:Figure2}C), respectively. The predictions of two models are overlaid on preoperative (Figure 2D: baseline, Figure \ref{Fig:Figure2}F: EMReDL) and recurrence (Figure \ref{Fig:Figure2}E: baseline, Figure \ref{Fig:Figure2}G: EMReDL) T1C images. Note the recurrence area is well beyond the contrast-enhancing core tumor on the preoperative MRI, which showed higher correspondence with the infiltrated area identified by EMReDL compared to baseline model. This improvement could possibly be explained by the tumor invasion area revealed by the physiological MRI shown underneath. Note the ground truth (green line) of the whole tumor region was taken as the combination of the core tumor and the recurrence tumor, with the assumption that the infiltrated tumor in the FLAIR is more responsible for the recurrence outside of the core tumor than other regions. 

\graphicspath{ {./images/} }

Moreover, the proposed method is compared with other CNN-based WSS models \citep{Kervadec2019,Roth2019,tang2018normalized} and SVM-based recurrence prediction model \citep{Akbari2016}. Table \ref{tab:Table2} showed that CNN-based WSS model achieved better accuracy than the voxel-wise SVM calssifier, suggesting the usefulness of CNN in capturing  the spatial information. Further, the EMReDL obtained higher accuracy than other WSS models, which supports the value of incorporating the physiological information through the physiological prior predictive model. As mentioned, physiological MRI has higher specificity in reflecting tumor biology but lower resolution than structural MRI. Benefiting from the prior predictive model, the physiological information could be effectively employed and less affected by the structural MRI, which hence could improve the model performance. In comparison, the pseudo labels generated through the normalized cut regularization in \citep{tang2018normalized} and the random walker regularization in \citep{Roth2019} were obtained by treating the structural and physiological MRI equally, therefore may not effectively leverage the information from physiological MRI. 

\begin{table}[!h] 
\caption{Comparison of weakly supervised models}
\label{tab:Table2}
{\centering
\begin{tabular}{ccccccc}
\hline
\multicolumn{2}{c}{} & \begin{tabular}[c]{@{}c@{}}Comparison \\ model   1\end{tabular} & \begin{tabular}[c]{@{}c@{}}Comparison \\ model   2\end{tabular} & \begin{tabular}[c]{@{}c@{}}Comparison \\ model   3\end{tabular} & \begin{tabular}[c]{@{}c@{}}Comparison \\ model   4\end{tabular} & EMReDL \\ \hline
\multirow{2}{*}{AUC} & Train & 0.764 & 0.901 & 0.855 & 0.923 & 0.971 \\
 & Test & 0.788 & 0.888 & 0.866 & 0.919 & 0.965 \\ \hline
\multirow{2}{*}{Sensitivity} & Train & 0.757 & 0.790 & 0.845 & 0.838 & 0.906 \\
 & Test & 0.765 & 0.764 & 0.824 & 0.815 & 0.898 \\ \hline
\multirow{2}{*}{Specificity} & Train & 0.664 & 0.934 & 0.799 & 0.882 & 0.918 \\
 & Test & 0.679 & 0.930 & 0.841 & 0.891 & 0.916 \\ \hline
\multirow{2}{*}{\begin{tabular}[c]{@{}c@{}}Youden\\    index\end{tabular}} & Train & 0.422 & 0.724 & 0.644 & 0.720 & 0.825 \\
 & Test & 0.444 & 0.693 & 0.664 & 0.706 & 0.813 \\ \hline
\multirow{2}{*}{MCC} & Train & 0.423 & 0.722 & 0.645 & 0.717 & 0.823 \\
 & Test & 0.444 & 0.685 & 0.658 & 0.697 & 0.808 \\ \hline
\multirow{2}{*}{Dice} & Train & 0.593 & 0.749 & 0.725 & 0.764 & 0.849 \\
 & Test & 0.621 & 0.727 & 0.739 & 0.755 & 0.846 \\ \hline
\end{tabular}
\par}
{AUC: area under the curve; MCC: Matthews correlation coefficient; Comparison model 1: SVM classifier based recurrence prediction; Comparison model 2: Normalized cut loss based WSS; Comparison model 3: Size-constrained loss based WSS; Comparison model 4: Random walker regularization based WSS.}
\end{table}

\begin{figure}[H]
\centering
 \includegraphics[scale=.29]{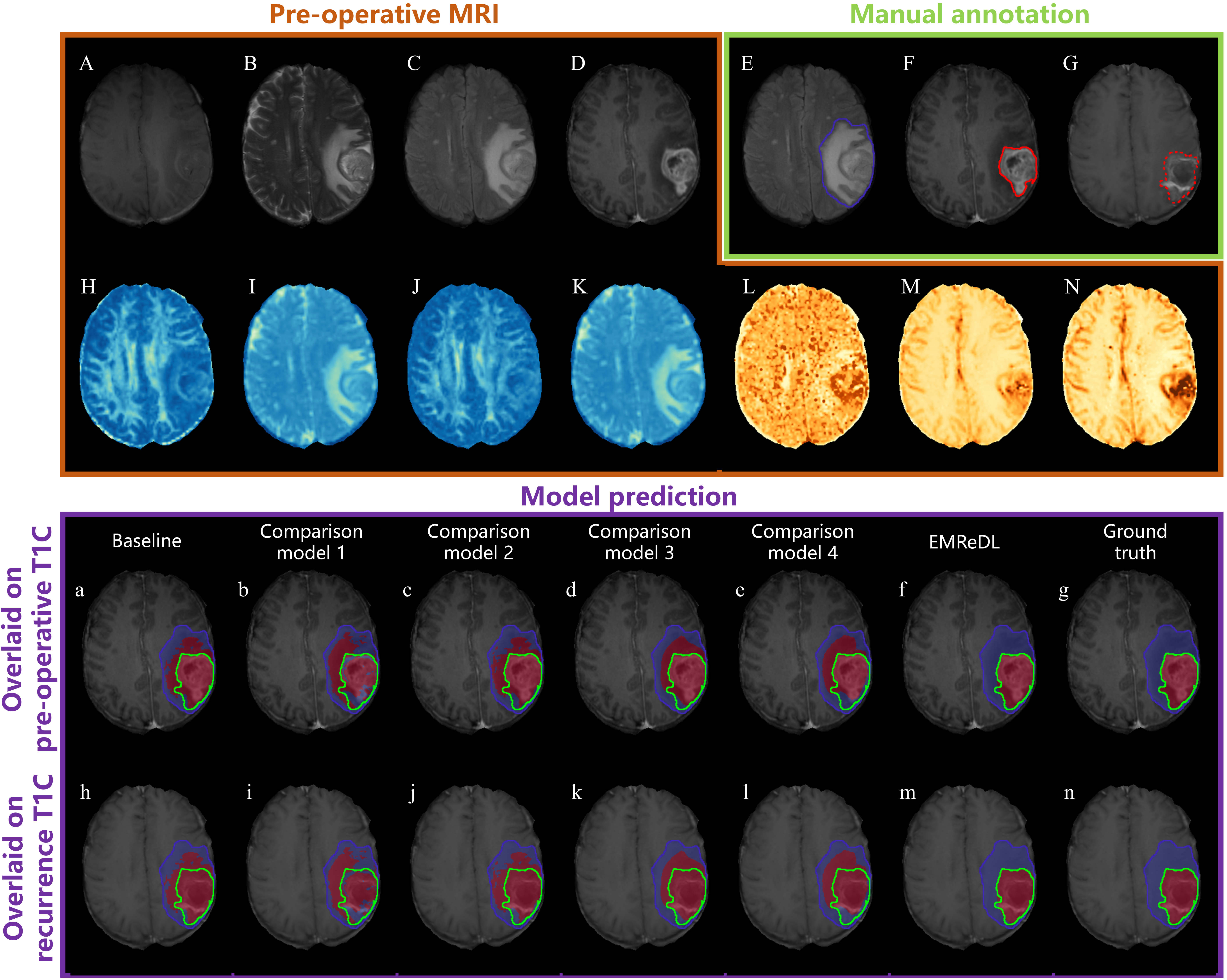}
\caption{Typical result of tumor infiltration region prediction by different models. Top panel: pre-operative and recurrence images. A$\sim$D: pre-operative T1, T2, FLAIR, T1C in sequence. E$\sim$G: labelled pre-operative FLAIR, T1C, and recurrence T1C; H$\sim$N: pre-operative FA, MD, DTI-q, DTI-p, MTT, rCBF and rCBV images in sequence. Bottom panel. Predicted infiltrated region (red) with the ROI2 (blue) overlaid. The green lines indicate the ground truth. a$\sim$g: segmentation of different models overlaid on pre-operative T1C images. h$\sim$n: segmentation of different models overlaid on recurrence T1C images. }
\label{Fig:Figure3}
\end{figure}

Figure \ref{Fig:Figure3} presents an typical example comparing the prediction of different models. Figure \ref{Fig:Figure3}A$\sim$D show the structural images. Figure \ref{Fig:Figure3}E and \ref{Fig:Figure3}F show the FLAIR abnormality and contrast-enhancing core tumor respectively, while Figure \ref{Fig:Figure3}G indicates the recurrence regions on the follow up scans. The physiological MRI are shown in Figure 2H$\sim$N. Compared with previous methods, the proposed method obtained segmentation result more similar to reference segmentation, especially in predicting the infiltrated region beyond contrast-enhanced core tumor.

Moreover, quantitative analysis is given in Table \ref{tab:Table3} for comparing different models in segmenting infiltrated tumor region beyond enhanced core tumor. As expected, all models obtained lower performance than segmenting the whole tumor region including the core tumor, which may be because we take the recurrence region as the reference infiltration region, while some non-recurrence area may also display invasive imaging features in the preoperative MRI. Additionally, the EMReDL achieved higher performance than other compared models in five different evaluation metrics, which imply the feasibility of the constructed regularizer. 
\begin{table}[!h]
{\centering
\caption{Comparison of infiltrated tumor segmentation}
\label{tab:Table3}
\resizebox{\textwidth}{!}{%
\begin{tabular}{cccccccc}
\hline
 &  & \begin{tabular}[c]{@{}c@{}}Comparison   \\ model 1\end{tabular} & \begin{tabular}[c]{@{}c@{}}Comparison   \\ model 2\end{tabular} & \begin{tabular}[c]{@{}c@{}}Comparison   \\ model 3\end{tabular} & \begin{tabular}[c]{@{}c@{}}Comparison   \\ model 4\end{tabular} & EMReDL \\ \hline
\multirow{2}{*}{AUC} & Train & 0.781 & 0.680 & 0.778 & 0.804 & 0.915 \\
 & Test & 0.807 & 0.701 & 0.807 & 0.837 & 0.938 \\ \hline
\multirow{2}{*}{Sensitivity} & Train & 0.787 & 0.480 & 0.771 & 0.736 & 0.809 \\
 & Test & 0.801 & 0.517 & 0.790 & 0.779 & 0.876 \\ \hline
\multirow{2}{*}{Specificity} & Train & 0.664 & 0.860 & 0.676 & 0.757 & 0.890 \\
 & Test & 0.679 & 0.858 & 0.711 & 0.774 & 0.889 \\ \hline
\multirow{2}{*}{\begin{tabular}[c]{@{}c@{}}Youden   \\ index\end{tabular}} & Train & 0.451 & 0.340 & 0.448 & 0.493 & 0.699 \\
 & Test & 0.480 & 0.375 & 0.501 & 0.553 & 0.765 \\ \hline
\multirow{2}{*}{MCC} & Train & 0.400 & 0.356 & 0.398 & 0.450 & 0.677 \\
 & Test & 0.449 & 0.398 & 0.471 & 0.527 & 0.746 \\ \hline
 \multirow{2}{*}{Dice} & Train & 0.408 & 0.346 & 0.407 & 0.441 & 0.621 \\
 & Test & 0.478 & 0.398 & 0.492 & 0.528 & 0.711 \\ \hline
\end{tabular}}
{AUC: area under the curve; MCC: Matthews correlation coefficient; Comparison model 1: SVM classifier based recurrence prediction; Comparison model 2: Normalized cut loss based WSS; Comparison model 3: Size-constrained loss based WSS; Comparison model 4: Random walker regularization based WSS.}
}
\end{table}

To summarize, the model comparisons may validate the performance of the proposed weakly supervised model. Also, our model showed comparable performance in both training and testing sets, which could suggest the robustness of the model. 

\graphicspath{ {./images/} }

\subsection{MRS results}
The MRS results showed that the predicted infiltrated region showed significantly more aggressive signature than the non-infiltrated region, which suggests the infiltration prediction could have significance regarding the tumor-induced metabolic change.  Specifically, choline is a marker of cellular turnover and membrane integrity, which is correlated with tumor proliferation. NAA is a maker of neuron structure, which may be destructed by the tumor infiltration. In previous studies, the choline/NAA ratio was frequently used an imaging marker to indicate tumor invasiveness, which was shown to correlate with patient outcomes. The detailed comparison of MRS data from the predicted infiltrated ad non-inlfiltrated regions are detailed in Table \ref{tab:Table6}.

\graphicspath{ {./images/} }
\begin{figure}[!h]
\centering
\includegraphics[scale=.38]{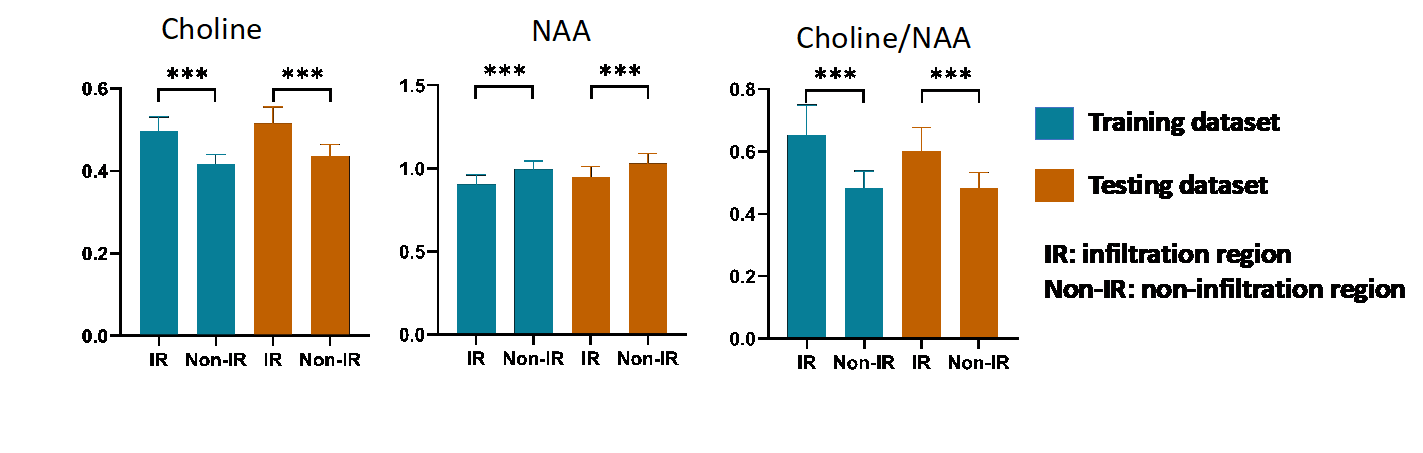}
\caption{MRS comparison of the infiltrated and non-infiltrated regions}
\label{Fig:Figure4}
\end{figure}

\begin{table}[!h]
\caption{MRS comparison of the segmented infiltration and non-infiltration }
\label{tab:Table6}
\centering{
\begin{tabular}{ccccc}
\hline
\multicolumn{2}{c}{  } & IR & Non-IR & p-value \\ \hline
\multirow{2}{*}{Choline} & Training & 0.50±0.13 & 0.42±0.09 & 3.1×$10^{-12}$ \\
 & Testing & 0.52±0.14 & 0.44±0.11 & 4.0×$10^{-10}$ \\ \hline
\multirow{2}{*}{Cho/NAA} & Training & 0.65±0.35 & 0.48±0.20 & 1.4×$10^{-8}$ \\
 & Testing & 0.60±0.27 & 0.48±0.18 & 4.1×$10^{-7}$ \\ \hline
\multirow{2}{*}{NAA} & Training & 0.90±0.22 & 0.99±0.20 & 9.3×$10^{-8}$ \\
 & Testing & 0.95±0.24 & 1.03±0.21 & 5.9×$10^{-6}$ \\ \hline
\end{tabular}
\par}
{IR: infiltration region; NAA: N-acetylaspartate}
\end{table}

Our study has limitations. Firstly, our manual labels were delineated by human experts. Therefore, different from the synthetic images, any analysis performed on this dataset may be biased and subjective compared to the synthetic images. Secondly, the other weakly supervised models that we compared with our models are not developed based on MRI. Therefore the performance may be affected when applied to our images. Lastly, due to the nature of tumor infiltration and ethics issue, some infiltrated tumor may not be directed observed and measured, as some tumor regions are more sensitive to treatment, Therefore, incorporating longitudinal MRI into the model could yield a more accurate infiltrated tumor estimation, which we are improving in our current study. 

\section{Conclusions}
In this paper, we presented an expectation-maximization regularized weakly supervised tumor segmentation model based on the deep convolutional neural networks. The proposed method was developed to segment both the core and peritumoral infiltrated tumor based on the multiparametric MRI. This weakly supervised model was developed to tackle the challenge of obtaining the full accurate labels for the infiltrated tumour. To effectively leverage the physiological MRI that has higher specificity but lower resolution than structural MRI, we constructed a physiological prior map generated from a fully connected neural network, for the iterative optimization of the CNN segmentation model. Using the tumor burden, tumor recurrence and MRS, the model evaluation confirms that our proposed model achieved higher  accuracy than the published state-of-the-art weakly supervised methods, using the regularizer constructed from physiological MRI.

\bibliography{references}

\section{Appendix}

\subsection{Implementation of comparison methods}
\label{sec:ImpleComp}
In this study, the EMReDL was compared with five different models: U-net with PCE loss, voxel-wise SVM classifier based recurrence prediction, and three state-of-the-art weakly supervised segmentation (WSS) methods using voxel-level partial annotations. The U-net with PCE is trained with the same setting as EMReDL, except that the proposed regularized term $J_{\text{reg}}(\mtheta)$ is removed from the loss function.

For recurrence prediction based on SVM classifier \citep{Akbari2016}, the voxel-wise multi-parametric MR signature was used as the input of the SVM classifier, and two extremities were selected to train the model based on the expectation that are likely to have relatively lower and higher infiltration. The distal edge of edema was defined as the far extremity, while the area closely adjacent to the enhancing core tumor was designated as the near extremity \citep{Akbari2016}. In our experiment, this model was tested by selecting the voxels within 2 mm outside the core tumor mask as the near extremity, and selecting the voxels within 2 mm inside the edema region as the far extremity. The rule and parameter setting followed the design of the original implementation \citep{Akbari2016}.

For normalized cut loss based WSS \citep{tang2018normalized}, the original model (https://github.com/meng-tang/rloss) was adapted by replacing the RGB input image with our multiparametric MR images, and treating the manually labelled ROI1 and ROI3 as the scribble annotation.

For size-constrained  loss  based  WSS \citep{Kervadec2019}, the best performed model (PCE combined with size loss and tags loss) was adopted and tested. The size loss enforces the sum over all probabilities in the probability map is greater than lower bound, and less than the upper bound on the object size. In our experiment, we set the size of core tumor (ROI1) as lower bound on tumor size, and the summarized size of tumor core and edema (ROI1\&ROI2) as the upper bound on tumor size. Apart from the input layer adaptation, other experiment conditions were kept the same as the original implementation (https://github.com/LIVIAETS/SizeLoss\_WSS).

For random walker regularization based WSS \citep{Roth2019}, the initial segmentation was generated from partial labels using random walker algorithm, which is considered as the pseudo training label for fully convolutional network (FCN). Afterwards, the training step and regularization step were iteratively performed for learning the segmentation model. For the regularization step, the region around the segmentation boundaries $\left( P(\mbx)\approx0.5 \right)$ from FCN prediction is considered as uncertain area, which was relabeled by random walker algorithm to produce the pseudo training label for next training iteration. This method was implemented according to its original hyperparameters and training schemes \citep{Roth2019}.


\small

\end{document}